\documentstyle[twoside,fleqn,espcrc2]{article}

\newcommand{\nc}{\newcommand}
\nc{\jv}{\frac{J}{4}}
\nc{\neel}{Ne\'{e}l}
\nc{\qq}{\P}
\nc{\rng}{\rangle}
\nc{\lng}{\langle}
\nc{\rcite}{ref.\ \cite}
\nc{\ba}{\begin{array}}
\nc{\ea}{\end{array}}
\nc{\Id}{I}
\nc{\Nul}{{\bf 0}}
\nc{\reals}{{\sf I}\kern-.12em{\sf R}}
\nc{\compl}{{\sf I}\kern-.48em{\sf C}}
\nc{\Z}{{\sf Z}}
\nc{\lb}{\left(}
\nc{\rb}{\right)}
\nc{\qrt}{\frac{1}{4}}

\newcommand{\gm}{\gamma}
\newcommand{\dl}{\delta}

\newcommand{\noi}{\noindent}
\newcommand{\half}{\frac{1}{2}}
\newcommand{\hhalf}{{\textstyle \frac{1}{2}}}

\newcommand{\rf}[1]{(\ref{#1})}

\newcommand{\ra}{\rightarrow}
\newcommand{\be}{\begin{equation}}
\newcommand{\ee}{\end{equation}}
\newcommand{\bea}{\begin{eqnarray}}
\newcommand{\eea}{\end{eqnarray}}
\nc{\nn}{\nonumber}
\nc{\thpm}{'t~Hooft-Polyakov monopole}
\nc{\eqrf}{eq.\ \rf}
\nc{\site}{x}
\nc{\shat}{\hat{S}}
\nc{\Tr}{\mbox{Tr}\;}
\nc{\defeq}{\stackrel{{\rm def}}{=}}
\nc{\calo}{{\cal O}}

\newcommand{\plb}[1]{Phys.~Lett.~#1B\ }

\newcommand{\npb}[1]{Nucl.~Phys.~B#1\ }
\newcommand{\nproc}[1]{Nucl.\ Phys.~B (Proc.~Suppl.)~#1\ }

\nc{\ACK}{\vspace{2ex} \noi {\bf Acknowledgement\\[1ex]}}

\newcommand{\AmS}{{\protect\the\textfont2
  A\kern-.1667em\lower.5ex\hbox{M}\kern-.125emS}}

\title{Gauge invariant extremization\thanks{Oxford preprint
OUTP-92-25P; hep-lat/9210038}}

\author{A.J.\ van der Sijs\address{Theoretical Physics, University of
Oxford, 1 Keble Road, Oxford  OX1 3NP, United Kingdom \\
    {[e-mail: {\protect\tt vdsijs@dionysos.thphys.ox.ac.uk}]}
}\thanks{Supported by SERC grant GR/H01243.}}
\begin{document}

\begin{abstract}
Recently, Duncan and Mawhinney introduced a method to
find saddle points of the action in simulations
of non-abelian lattice gauge theory.
The idea, called `extremization', is to minimize
$\int(\dl S/\dl A_\mu)^2$ instead of the action $S$ itself as in
conventional `cooling'.
The method was implemented in an explicitly gauge variant way,
however, and gauge dependence showed up in the results.

Here we present a gauge invariant formulaton of extremization on the
lattice, applicable to
any gauge group and any lattice action. The procedure is
worked out in detail for
U(1) and SU(N) lattice gauge theory with the plaquette action.
\end{abstract}

\maketitle

\section{INTRODUCTION}
A (lattice) quantum field
configuration may be considered as a (semi)classical background with
ultraviolet fluctuations superimposed. The background may contain
long distance objects such as magnetic monopoles and instantons
in addition to the classical vacuum.
It would be interesting to make this semiclassical background
visible by somehow stripping off the short distance fluctuations.

The cooling method \cite{cooling} is aimed at removing ultraviolet
fluctuations for a field theory on the lattice.
Starting from a Monte Carlo configuration the action $S$ is lowered
by making small (local) changes in the fields, which remove
short distance fluctuations.
The configuration is iteratively smoothened and evolved towards a
local minimum of the action.
It is important to realize, however, that the background itself is also
affected in the process. The cooling algorithm would move a
configuration away from an unstable classical soluton, for example.

It would therefore be interesting to have an algorithm capable of
reaching any classical solution, not just local minima. Then one might
also find saddle points, stationary points with one or more
unstable directions.

The sphaleron in the electroweak sector of the standard model is an
example of a saddle point configuration which has received a lot of
attention recently in the context of baryon number violating processes.
Another example of a saddle point is provided by mutually attracting
particles in a periodic volume.
On a circle, for example, the configuration in which two such particles
are at antipodal points is an unstable stationary point.
This kind of saddle point is of particular interest for studies of
monopoles and instantons in gauge theories on a lattice with periodic
boundary conditions.

\subsection{Extremization}

Duncan and Mawhinney \cite{dunmaw1} recently proposed a method to find saddle
points of the action in lattice field configurations. This method, called
`extremization', is based on lowering the `extremization action' $\shat$,
\be
\shat = \sum_{\site,\mu,a} \lb \frac{\dl
S}{\dl A_{\mu}^a (\site)} \rb^2  \; , \label{101}
\ee
instead of the action $S$ itself.
It is clear that $\shat$ attains its minimal value of zero
for any solution to the equations of motion.
Minimization of $\shat$ will move a configuration towards
the classical solution in whose basin of attraction
it lies.

In ref.~\cite{dunmaw1} the extremization idea was
implemented in a gauge variant way.
As a consequence, extremization of initial configurations differing by a
gauge transformation led to different, gauge non-equivalent,
final configurations.

The occurrence and location of saddle points is determined
by the equations of motion, however, and should not depend
on the gauge choice.
Indeed, $\shat$ is just the integral
over space-time of the equations of motion squared
and the extremization action density is gauge invariant.
For the (continuum) Yang-Mills action, for example,
\be
\shat^{\mbox{{\scriptsize cont}}} = \int_V (D_{\mu}
F_{\mu \nu}^a)^2 \; . \label{102}
\ee
Hence it must be possible to define the extremization
procedure in a gauge independent manner. This should remain true
for the lattice regulated theory, since the lattice preserves
gauge symmetries.

Here we formulate a gauge invariant extremization procedure (see also
ref.~\cite{sijs}). It is shown that the extremization action $\shat$
on the lattice can be defined unambiguously,
for any gauge group and lattice action.
We focus on U(1) and SU(N) gauge theory with
the plaquette action to illustrate the method.

\section{GAUGE INVARIANT EXTREMIZATION ON THE LATTICE}
How do we define the extremization action $\shat$ on the lattice?
The basic idea is simple. The lattice action $S$ is a functional
of the link variables $U_\mu(x)$, so it is natural to express $\shat$ in
terms of derivatives of $S$ with respect to these variables. The
field $A_\mu(x)$, defined as the logarithm of the link variables,
for example, need not be introduced at all.
The lattice extremization action will look like
\be
\shat \propto
\sum_{\stackrel{\rm links}{U}}\;\;
\raisebox{1ex}{``} \;\lb\frac{\dl S}{\dl U}\rb^2 \;\raisebox{1ex}{''} \; .
\label{200}
\ee
In the following we will make this more precise.

\subsection{Extremization action on the lattice}

Consider the plaquette action for SU(N) gauge theory,
\be
S =  - \frac{1}{g^2} \sum_\Box \lb \Tr U_\Box + \Tr U_\Box^\dagger
\rb \; . \label{201}
\ee
Focusing attention on a particular link $U$,
this can be written as
\bea
S &=& - \frac{1}{g^2} \;
\Tr \lb U F_U + U^\dagger F_U^\dagger \rb \label{202}\\
&&\mbox{+ terms independent of $U$} \; . \nn
\eea
Here $F_U$ is the `force' for the link $U$, the
sum of its $\gm$ `staples',
\thicklines
\setlength{\unitlength}{1.5ex}
\be
F_{U_\mu(x)} = \sum_{\nu\neq\mu} \;\;
\lb
\raisebox{-\unitlength}{
\begin{picture}(6,6)(0,4)
\put(1,5){\circle*{0.4}}
\put(5,5){\circle*{0.4}}
\put(5,5){\vector(0,1){2}}
\put(5,7){\line(0,1){2}}
\put(5,9){\vector(-1,0){2}}
\put(3,9){\line(-1,0){2}}
\put(1,9){\vector(0,-1){2}}
\put(1,7){\line(0,-1){2}}
\put(0.5,3.5){\mbox{$x$}}
\end{picture}
}
 +
\raisebox{-\unitlength}{
\begin{picture}(6,6)(0,4)
\put(1,5){\circle*{0.4}}
\put(5,5){\circle*{0.4}}
\put(5,5){\vector(0,-1){2}}
\put(5,3){\line(0,-1){2}}
\put(5,1){\vector(-1,0){2}}
\put(3,1){\line(-1,0){2}}
\put(1,1){\vector(0,1){2}}
\put(1,3){\line(0,1){2}}
\put(0.5,6){\mbox{$x$}}
\end{picture}
}
\rb
 \; , \label{203}
\ee
where
\be
\gm = 2\,(d-1)  \label{299}
\ee
in $d$ dimensions.
Note that $F=F_U$ is not an SU(N) matrix in general, but if the gauge
group is SU(2) it can always be written as a constant times an SU(2) matrix.

 From expression \rf{202} one can read off the derivatives of $S$
with respect to the link $U=U_\mu(\site)$.
A variation $\dl U$ in the link
gives rise to a change in the action
\bea
\dl S &=& - \Tr \left\{ \lb \frac{\dl S}{\dl U} - U^{\dagger}
      \frac{\dl S}{\dl U^{\dagger}} U^{\dagger}\rb \dl U \right\}
      \label{102b} \\
      &=& - \Tr \left\{ \lb F - U^{\dagger}
      F^{\dagger} U^{\dagger}\rb \dl U \right\}
      \label{102b2} \\
&=& - \Tr \left\{ (UF-(UF)^\dagger)\, \dl U\, U^\dagger \right\}
     \; . \label{102c}
\eea
Factors of $1/g^2$ are dropped from now on and we have used that
\be
0 = \dl (U^{\dagger}U) = (\dl U^\dagger)U + U^\dagger \dl U
 \; .  \label{102f}
\ee

Pictorially, the `coefficient' of $(\dl U \, U^\dagger)$ in \eqrf{102c}
looks like
\be
UF-(UF)^\dagger =
\sum
\;\lb
\raisebox{-1.6\unitlength}{
\begin{picture}(5,5)(0.5,0.5)
\put(1.5,1){\circle*{0.4}}
\put(1,1.5){\circle*{0.4}}
\put(0.5,-0.5){\mbox{$x$}}
\put(1.5,1){\vector(1,0){2}}
\put(3.5,1){\line(1,0){1.5}}
\put(5,1){\vector(0,1){2}}
\put(5,3){\line(0,1){2}}
\put(5,5){\vector(-1,0){2}}
\put(3,5){\line(-1,0){2}}
\put(1,5){\vector(0,-1){2}}
\put(1,3){\line(0,-1){1.5}}
\end{picture}
}
-
\raisebox{-1.6\unitlength}{
\begin{picture}(5,5)(0.5,0.5)
\put(1.5,1){\circle*{0.4}}
\put(1,1.5){\circle*{0.4}}
\put(0.5,-0.5){\mbox{$x$}}
\put(1,1.5){\vector(0,1){2}}
\put(1,3.5){\line(0,1){1.5}}
\put(1,5){\vector(1,0){2}}
\put(3,5){\line(1,0){2}}
\put(5,5){\vector(0,-1){2}}
\put(5,3){\line(0,-1){2}}
\put(5,1){\vector(-1,0){2}}
\put(3,1){\line(-1,0){1.5}}
\end{picture}
}
\rb
\label{102ga}
\ee
where the sum runs over the $\gm$ staple directions.
Here the diagrams have not been closed at $x$
in order to indicate that
they represent non-traced products of matrices along lines
starting and terminating at $x$.

The lattice extremization action $\shat$ in
\eqrf{200} can now be more precisely defined in terms of these
coefficients,
\bea
\lefteqn{\shat} \nn \\
 \lefteqn{\;\;= {\displaystyle \sum_{\stackrel{\rm links}{U}}} \half \Tr \{
(UF - (UF)^\dagger)^\dagger (UF-(UF)^\dagger)
\} } \label{104} \\
 \lefteqn{\;\;= {\displaystyle \sum_{\stackrel{\rm links}{U}}}  \half \Tr \{
2F^\dagger F - F U F U - (F U F U)^\dagger \} .}  \label{105}
\eea
Note that the same expression would have been obtained if the coefficients
of $\dl U$ in \eqrf{102b2} had been used.

$\shat$ is a sum over traces of closed Wilson loops. Thus, it is clearly
gauge invariant.

\subsection{A diagrammatical survey}

The various terms of $\shat$ coming from a link $U$
can be constructed by taking one of the open
diagrams from \eqrf{102ga} and connecting it to (the hermitian conjugate
of) another diagram.
In part of the diagrams so constructed the
link $U$ itself cancels out. This concerns the Wilson loops of length 6
contained in the $F^\dagger F$ term in \eqrf{105} and depicted in
fig.~$1a$ (here the dashed line is the link $U$). Some of these diagrams,
where the two staples constituting the loop lie on top of each other, are
trivial. Note that the diagram in fig.~$1b$, zigzagging around a cube,
does not occur.
\setlength{\unitlength}{3ex}
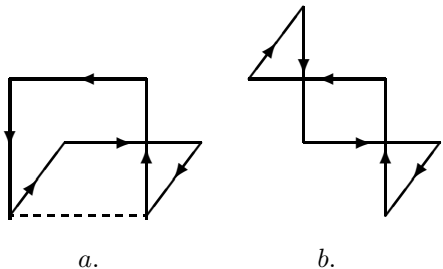
\begin{figure}[htb]
\begin{picture}(15,8)
\put(0,2){%
\raisebox{-\unitlength}{
\begin{picture}(6,7.5)(0,4)
\put(5,5){\vector(0,1){2}}
\put(5,7){\line(0,1){2}}
\put(5,9){\vector(-1,0){2}}
\put(3,9){\line(-1,0){2}}
\put(1,9){\vector(0,-1){2}}
\put(1,7){\line(0,-1){2}}
\put(1,5){\vector(3,4){0.8}}
\put(1.8,6.066667){\line(3,4){0.8}}
\put(2.6,7.133333){\vector(1,0){2}}
\put(4.6,7.133333){\line(1,0){2}}
\put(6.6,7.133333){\vector(-3,-4){0.8}}
\put(5.8,6.066667){\line(-3,-4){0.8}}
\put(1,5){\dashbox{0.2}(4,0){}}
\put(3,3.5){\mbox{$a.$}}
\end{picture}
}
}
\put(7,2){%
\raisebox{-\unitlength}{
\begin{picture}(6,7.5)(0,4)
\put(5,5){\vector(0,1){2}}
\put(5,7){\line(0,1){2}}
\put(5,9){\vector(-1,0){2}}
\put(3,9){\line(-1,0){2}}
\put(1,9){\vector(3,4){0.8}}
\put(1.8,10.066667){\line(3,4){0.8}}
\put(2.6,11.133333){\vector(0,-1){2}}
\put(2.6,9.133333){\line(0,-1){2}}
\put(2.6,7.133333){\vector(1,0){2}}
\put(4.6,7.133333){\line(1,0){2}}
\put(6.6,7.133333){\vector(-3,-4){0.8}}
\put(5.8,6.066667){\line(-3,-4){0.8}}
\put(3,3.5){\mbox{$b.$}}
\end{picture}
}
}
\end{picture}

\caption{
$\mbox{}$\protect\\
$a$. Length 6 Wilson loop contributing to $\shat$.\protect\\
$b$. Non-contributing loop of length 6.}
\end{figure}

The other diagrams in the $U$-contribution to $\shat$,
corresponding to the $FUFU$ terms in
\eqrf{105}, contain the link $U$ twice. They are closed loops of length
8, see fig.~$2a$, and among them are the `double plaquettes' shown in
fig.~$2b$.
\begin{figure}[htb]
\begin{picture}(15,8)
\put(0,2){%
\raisebox{-\unitlength}{
\begin{picture}(6,7.5)(0,4)
\put(4.7,5){\vector(0,1){2}}
\put(4.7,7){\line(0,1){1.6}}
\put(4.7,8.6){\vector(-1,0){2}}
\put(2.7,8.6){\line(-1,0){2}}
\put(0.7,8.6){\vector(0,-1){2}}
\put(0.7,6.6){\line(0,-1){2}}
\put(0.7,4.6){\vector(1,0){2}}
\put(2.7,4.6){\line(1,0){2}}
\put(4.7,4.6){\vector(3,4){1.1}}
\put(5.8,6.066667){\line(3,4){0.8}}
\put(6.6,7.133333){\vector(-1,0){2.5}}
\put(4.1,7.133333){\line(-1,0){1.5}}
\put(2.6,7.133333){\vector(-3,-4){0.8}}
\put(1.8,6.066667){\line(-3,-4){0.8}}
\put(1,5){\vector(1,0){2}}
\put(3,5){\line(1,0){1.7}}
\put(3,3.5){\mbox{$a.$}}
\end{picture}
}
}
\put(7,2){%
\raisebox{-\unitlength}{
\begin{picture}(6,7.5)(0,4)
\put(5,4.6){\vector(0,1){2}}
\put(5,6.6){\line(0,1){2.4}}
\put(5,9){\vector(-1,0){2}}
\put(3,9){\line(-1,0){2}}
\put(1,9){\vector(0,-1){2}}
\put(1,7){\line(0,-1){2}}
\put(1,5){\vector(1,0){2}}
\put(3,5){\line(1,0){1.7}}
\put(4.7,5){\vector(0,1){2}}
\put(4.7,7){\line(0,1){1.6}}
\put(4.7,8.6){\vector(-1,0){2}}
\put(2.7,8.6){\line(-1,0){2}}
\put(0.7,8.6){\vector(0,-1){2}}
\put(0.7,6.6){\line(0,-1){2}}
\put(0.7,4.6){\vector(1,0){2}}
\put(2.7,4.6){\line(1,0){2.3}}
\put(3,3.5){\mbox{$b.$}}
\end{picture}
}
}
\end{picture}

\caption{
$\mbox{}$\protect\\
$a$. Contributing diagram of length 8.\protect\\
$b$. `Double plaquette'.}
\end{figure}

The number of terms occurring in $S$, \eqrf{104}, is
$(2\gm)^2\, N_{{\rm link}}$,
but the number of independent Wilson loops is much smaller.
This is because all the diagrams occur at least twice, each of them
is accompanied by its hermitian conjugate, and some of them are trivial.
To be precise, there are $\half \gm (\gm-1)\, N_{{\rm link}}$
different loops of length 6 and
$\half\gm(\gm-\half)\, N_{{\rm link}}$ of length 8.
(Note that the double
plaquettes are shared among the 4 links contributing to it.) This gives a
total number of $[\qrt\gm(4\gm - 3)]\,N_{{\rm link}}$
different non-trivial diagrams.
The number between the square brackets equals $5/2$, $13$, $63/2$ in
2, 3 and 4 dimensions, respectively.

If one only wants to calculate $\shat$ for a particular configuration one
need not compute all these diagrams separately. It is easier to
compute $UF-(UF)^\dagger$ and square it.
Minimization of $\shat$, needed for extremization of $S$,
is more complicated however. In order to perform
a local update of $\shat$ at $U$ one has to consider all the diagrams
depending on $U$.
Since there are diagrams quadratic as well as linear in $U$, it is not
straightforward to find the minimum of $\shat$ with respect to $U$.
An iterative algorithm may be required.

Let us compute the number of diagrams which has to be taken into account
in such an update step.
There are $6\gm(\gm-1)$ different diagrams depending linearly on $U$,
half of them of length 6 and half of length 8.
The number of diagrams depending quadratically on $U$
is $\half\gm(\gm+1)$. These are all of length 8.
This adds up to a total number of
$\half\gm(13\gm-11)$ different diagrams depending on $U$.
For $d=2, 3, 4$, this becomes 15, 82 and 201, respectively.

\subsection{The limit $a\ra 0$}

For smooth fields the plaquette action reduces to the Yang-Mills action
in leading order in the lattice spacing $a$,
and one expects the
lattice extremization action $\shat$ to reduce to the continuum
extremization action in \eqrf{102} accordingly.

It is easy to check this in the case of the U(1) theory, for which
\be
S_{U(1)} = - \frac{1}{2g^2} \, \sum_\Box \, (\Box + \Box^*)
\label{107}
\ee
and
\be
\shat_{U(1)} = \qrt \sum_{\stackrel{\rm links}{U}}
(UF - (UF)^*)^* (UF-(UF)^*)
   . \label{109}
\ee
The expansion of the plaquette gives
\setlength{\unitlength}{1.5ex}
\bea
\lefteqn{
\raisebox{-1.5\unitlength}{
\begin{picture}(6,6)(0,0)
\put(1.5,1){\circle*{0.4}}
\put(1,1.5){\circle*{0.4}}
\put(0.5,-0.5){\mbox{$x$}}
\put(1.5,1){\vector(1,0){2}}
\put(3.5,1){\line(1,0){1.5}}
\put(5,1){\vector(0,1){2}}
\put(5,3){\line(0,1){2}}
\put(5,5){\vector(-1,0){2}}
\put(3,5){\line(-1,0){2}}
\put(1,5){\vector(0,-1){2}}
\put(1,3){\line(0,-1){1.5}}
\end{picture}
}
= }
\nn \\[1ex]
&&\exp \, [-ia^2 F_{\mu\nu}(x+\hhalf a \hat{\mu} + \hhalf a
    \hat{\nu}) + \calo(a^4)] , \label{210}
\eea
where terms of ${\cal O}(a^3)$ are absent because we expand around the
centre of the plaquette.
Using \eqrf{102ga} one deduces that
\bea
\lefteqn{UF - (UF)^* = } \nn \\
&&-2ia^3 \sum_\nu \partial_\nu F_{\mu\nu}(x+\hhalf a \hat{\mu})
+ \calo(a^5)  \; . \label{213}
\eea
Upon insertion in \eqrf{109} this gives
\be
\shat_{U(1)} = \int_V a^2 (\partial_\nu F_{\mu\nu} )^2 +\calo(a^4)
\label{214}
\ee
which is the abelian version of \eqrf{102}.
The extra factor of $a^2$ here comes from the
correspondence $\dl U \propto a\;\dl A$.

This result can be automatically generalized to the non-abelian case.
The same derivation applies if the derivatives are replaced by covariant
ones and the coefficients in $S$ and $\shat$ are
adjusted appropriately.

We end this section with two remarks about the extremization action
and its continuum limit \eqrf{102}.
Firstly, note that the continuum limit of the extremization action
is Lorentz invariant.
This can be contrasted
with tree level `improvement terms' \cite{improved}
for the plaquette action, whose leading $a$-behavior is of the
Lorentz non-invariant form
\be
\sum_{\mu,\nu} D_\nu F_{\nu\mu} \; D_\nu F_{\nu\mu} \label{298}
\ee
(with all the four subscripts $\nu$ equal).
Secondly, note that the construction of the lattice extremization action
provides one in a straightforward and elegant way with a lattice action
for the `higher derivative action' of \eqrf{102}.

\section{DISCUSSION}

The gauge invariant extremization procedure proposed here,
and illustrated for the lattice gauge theory with the plaquette action
for gauge groups U(1) and SU(2),
is applicable
to any lattice action and any gauge group. Discrete gauge groups, such
as the subgroups $\Z_N$ of U(1) and the non-abelian discrete subgroups
of SU(2), are not excluded.

Gauge invariant extremization
has not yet been tried out in simulations. Implementing it
should be straightforward, although the number of diagrams
per link is rather large, especially in four dimensions.
It would be interesting to see how the method performs in practice.

\end{document}